# Where Do Smart Contract Security Analyzers Fall Short?

Tamer Abdelaziz
NYU Abu Dhabi
tamer.m@nyu.edu

Salma Alsaghir
NYU Abu Dhabi
saa9125@nyu.edu

Karim Ali
NYU Abu Dhabi
karim.ali@nyu.edu

## Abstract

Smart contracts underpin high-value ecosystems such as decentralized finance (DeFi), yet recurring vulnerabilities continue to cause losses worth billions of dollars. Although numerous security analyzers that detect such flaws exist, real-world attacks remain frequent, raising the question of whether these tools are truly effective or simply under-used due to low developer trust. Prior benchmarks have evaluated analyzers on synthetic or vulnerable-only contract datasets, limiting their ability to measure false positives, false negatives, and usability factors that drive adoption.

To close this gap, we present a mixed-methods study that combines large-scale benchmarking with practitioner insights. We evaluate six widely used analyzers (i.e., Confuzzius, Dlva, Mythril, Osiris, Oyente, and Slither) on 653 real-world smart contracts that cover three high-impact vulnerability classes from the OWASP Smart Contract Top Ten (i.e., reentrancy, suicidal contract termination, and integer arithmetic errors). Our results show substantial variation in accuracy (F1 = 31.2–94.6%), high false-positive rates (up to 32.6%), and runtimes exceeding 700 seconds per contract. We then survey 150 professional developers and auditors to understand how they use and perceive these tools. Our findings reveal that excessive false positives, vague explanations, and long analysis times are the main barriers to trust and adoption in practice. By linking measurable performance gaps to developer perceptions, we provide concrete recommendations for improving the precision, explainability, and usability of smart-contract security analyzers.

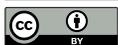

## 1 Introduction

Blockchain platforms such as Ethereum [10] have popularized smart contracts, which are self-executing programs that underpin decentralized applications. Critical ecosystems such as decentralized finance (DeFi) [60] and non-fungible token (NFT) marketplaces [59] now depend on them for transparent, verifiable transactions. This adoption has fueled the growth of the global blockchain industry, projected to reach US$ 393 billion by 2030 [38]. However, smart contracts remain frequent targets of security exploits. In 2024 alone, 303 attacks resulted in losses exceeding US$ 2.2 billion, an increase of 21% over the previous year [12]. From large-scale reentrancy exploits to subtle arithmetic errors, recurring vulnerabilities continue to erode developer and investor confidence [52, 53].

To mitigate these risks, the research community and industry have introduced a wide range of security analyzers that automatically detect vulnerabilities before deployment. These analyzers employ diverse techniques, including static analysis (e.g., Slither [24], Vandal [9]), symbolic execution (e.g., Mythril [39], Oyente [37]), fuzzing (e.g., Echidna [29], Confuzzius [57]), and more recently, learning-based methods (e.g., GPTScan [55], Dlva [3], SCooLS [2]). Despite the availability of several tools, real-world attacks continue to occur. This disconnect raises a central question: *are these security analyzers not sufficiently effective, or are developers not using them effectively (if at all)?*

Understanding this gap requires addressing two complementary facets. The first concerns the *technical effectiveness* of existing analyzers. In other words, how accurately and efficiently these tools detect vulnerabilities in real-world smart contracts. Although prior work (e.g., Ghaleb and Pattabiraman [26], Durieux et al. [20]) have established initial benchmarks, they often relied on synthetic datasets containing only vulnerable samples. This design prevents realistic estimation of false positive and false negative rates, which are key metrics to determine whether developers can trust and actively use these security analyzers. The second facet concerns *developer perception and trust.* Even technically sound tools would fail in practice, if users find them slow, imprecise, or unhelpful. Studies from other software domains [19, 33] show that developers frequently ignore analysis alerts that they do not perceive as actionable. However, little is known about whether smart contract developers face similar issues or what factors shape their trust in analyzer outputs.

To address these two facets, we conduct a *convergent parallel mixed-methods study* [54] that integrates quantitative benchmarking with qualitative insights from practitioners. On the quantitative side, we systematically evaluate six widely used analyzers (Confuzzius, Dlva, Mythril, Osiris, Oyente, and Slither) across 653 real-world smart contracts that include both vulnerable and benign code. To ensure a fair comparison, we focus on three high-impact vulnerability types from the OWASP Smart Contract Top 10 [5]: *reentrancy*, *suicidal contract termination*, and *integer arithmetic errors*, which are detectable by all studied tools. Our quantitive benchmarking shows that no analyzer reliably detects all three vulnerability types. While CONFUZZIUS excels at detecting reentrancy, OSIRIS is the best at detecting integer arithmetic errors. Additionally, SLITHER and OYENTE trade some accuracy for faster scanning time. Across all tools, suicidal contract termination is the most challenging to detect (F1-scores below 60%).

On the qualitative side, we survey 150 professional smart contract developers and auditors to understand how they perceive, adopt, and trust these analyzers. Overall, the survey participants are more familiar with classical vulnerability types (e.g., reentrancy) than more recent threats (e.g., flash-loan attacks [6]). More than two-thirds of the participants expect analysis results within 10 minutes, because it is critical to make an immediate decision to deploy a contract or seek further auditing from a security expert prior to deployment. We then draw several insights from the survey responses that help us determine what design factors encourage broader integration of security analyzers into real development workflows. In particular, we identify three key trust barriers that hinder tool adoption: high false positive rates (70% of participants), lack of clear explanations (65% of participants), and limited integration with developers' workflows (71% of participants). Further analysis of the survey responses show that most participants prefer brief explanations that contain fix suggestions (82%) and precise bug locations (78%). Few participants indicate their desire for more details such as contextualized explanations (17%), real-world examples (17%), or actionable remediation steps (20%).

By connecting measurable tool performance with developer trust and adoption behaviors, our study advances the understanding of how security analyzers can evolve from technically promising prototypes into trusted components of secure smart-contract development. In particular, this paper makes the following contributions:

- We present a balanced benchmark of six widely used smart contract analyzers, revealing the trade-offs between detection accuracy, computational efficiency, and vulnerability coverage across real-world smart contracts.
- We report results from a survey of 150 developers and auditors, providing the first empirical account of how practitioners perceive these analyzers, what drives or hinders their trust, and which explanation styles that they find most actionable.
- We integrate findings from both strands to explain why high detection accuracy alone does not translate into widespread, confident adoption. We then offer concrete recommendations for aligning analyzer design with developer needs.

*Availability.* All data, benchmarks, and anonymized survey responses are available in our replication package [1].

## 2 Background
### 2.1 Smart Contracts
Smart contracts are self-executing programs deployed on blockchain networks such as Ethereum [10]. They enforce agreements without intermediaries and automatically execute transactions once predefined conditions are met. To execute a transaction, it first is stored in the `mempool`, a public staging area (pool) for all pending, unconfirmed transactions waiting to be selected by a validator [23].

Since the code and state of smart contracts are immutable after deployment, any vulnerability in a contract may have irreversible financial consequences, making automated security analysis a critical safeguard. Therefore, detecting malicious transactions in real-time (i.e., while they are still in the `mempool`) is essential for preventing exploitation before the transactions are confirmed on the blockchain.

### 2.2 Smart Contract Vulnerabilities
Despite years of research and auditing practice, vulnerabilities in smart contracts remain both common and costly. Chainalysis [12] reports that in 2024 alone, security exploits across major blockchains caused losses exceeding US$2.2 billion. Independent audits continue to reveal that more than one-third of deployed contracts exhibit at least one known weakness [20]. These recurring failures highlight gaps between vulnerability detection tools and developer practices.

Although dozens of smart-contract vulnerability types have been catalogued, not all are equally destructive or detectable by automated analyzers. The OWASP Smart Contract Top 10 [5] identifies the most impactful types. However, many tools advertise coverage for only a subset of them. To ensure a fair and reproducible comparison across analyzers, this study focuses on three high-impact and widely supported types: *reentrancy*, *suicidal contract termination*, and *integer arithmetic errors*. Our selection is based on four practical criteria: (i) *historical impact*, as each class has caused documented multi-million-dollar losses; (ii) *analyzer coverage*, which enables a meaningful comparison across all evaluated tools on the same vulnerability types; (iii) *labeling reliability*, as programmatic predicates and on-chain evidence let us construct reproducible ground truth; and (iv) *representative error families*, ensuring our set spans control-flow (reentrancy), lifecycle and access control (suicidal contracts), and arithmetic errors (integer overflow/underflow). We exclude emerging threats such as flash-loan and oracle-manipulation attacks because developer familiarity with these classes remains low (see Section 4) and most current analyzers provide little or no detection support. Chaliasos et al. [13] reach a similar conclusion, showing that 75% of 127 documented high-impact DeFi attacks fall outside the detection capabilities of evaluated tools. Including such vulnerabilities would result in near-zero detection rates, preventing meaningful comparison.

*2.2.1 Reentrancy.* A reentrancy vulnerability occurs when a contract sends funds to an external address before updating its internal state, allowing a malicious callee to recursively re-enter the vulnerable function and drain funds. The 2016 DAO incident exploited this flaw to steal 3.6 million ETH ($\approx$ US$60 million at the time), which required a controversial hard fork of Ethereum to recover those funds [52]. Although modern frameworks (e.g., OpenZeppelin's `ReentrancyGuard`) may help prevent it, reentrancy remains among the most frequently exploited weaknesses in decentralized finance (DeFi) platforms.

*2.2.2 Suicidal Contract Termination.* Also known as a suicide or selfdestruct vulnerability, this issue occurs when contract functions that can call `selfdestruct()` lack sufficient access controls, such as ownership checks or function modifiers [22]. This allows an unauthorized actor to destroy the contract. A prominent example is the 2017 Parity Wallet hack, where an attacker destroyed a crucial library contract, permanently freezing approximately 500,000 ETH (worth roughly US$150 million at the time). Although Ethereum Improvement Proposal (EIP) 6049 [48] has deprecated the selfdestruct() opcode in newer Solidity versions, similar kill-switch logic remains common in upgradable proxy contracts, making this a persistent security concern.

We consider suicidal contract termination as a specific, well-bounded instance of access-control failure. It features a simple, programmatic predicate for detection (unrestricted selfdestruct()), and approximately two-thirds of the analyzers in our study detect it. In contrast, general access-control vulnerabilities (e.g., complex authorization logic or role misconfigurations) are highly context-dependent and require manual specification for consistent labeling, which compromises reproducibility.

*2.2.3 Integer Arithmetic Errors.* Arithmetic overflows and underflows occur when operations exceed numeric bounds. These issues may result in unintended values [49], allowing attackers to create unauthorized tokens or bypass balance checks. In fact, such bugs have led to token-minting and balance-manipulation exploits, including the 2023 Poolz Finance attack [53]. While Solidity 0.8.0 has introduced automatic overflow checks, legacy contracts and libraries that still rely on unchecked arithmetic remain vulnerable. Therefore, security analyzers must accurately identify arithmetic issues across compiler versions, and notify developers to use safe arithmetic operations (e.g., using OpenZeppelin's SafeMath [43]). To ensure that our evaluation includes detectable instances of this class, we restrict the arithmetic-related portion of our benchmark to contracts compiled with pre-0.8 compilers.

## 3 Empirical Evaluation of Security Analyzers

To evaluate how effectively current analyzers detect and report smart-contract vulnerabilities, we conduct a quantitative benchmark on the three vulnerability types introduced in Section 2, focusing on the trade-offs that shape developer trust and tool adoption.

### 3.1 Dataset Collection

Existing public datasets of vulnerable smart contracts contain duplicates, inconsistent labels, and overly synthetic samples [17]. To enable fair comparison, we curate ScBench, a novel dataset of real-world contracts with verified ground-truth labels that supports measurement of both false positives and false negatives across the three target vulnerability classes.

*3.1.1 Data Sources.* We collect 717 smart contracts from academic datasets [3, 14, 27, 31] and security audit reports published by major firms such as Trail of Bits [41], ConsenSys Diligence [18], and OpenZeppelin [42]. We adopt the manually verified vulnerability labels from these sources as our ground truth.

Our selection process imposes two key requirements on each contract. First, the contract must contain at least one of the three vulnerability types under study. Second, the verified source code must be publicly available to enable our ground-truth validation. For each selected contract, we fetch the verified source code and its metadata (including compiler version and deployment block) from Etherscan [21] and obtain the runtime bytecode from Google BigQuery [28]. Source code availability is a primary filter, because several academic datasets provide only contract addresses and labels, without accompanying source code or bytecode. For example, the dataset from Chen et al. [14] contains 756 addresses, but we could successfully retrieve verified source code for only 309 of them. Therefore, our collection pipeline automatically excludes contracts with unverified source code.

Table 1: Distribution of smart contracts in ScBench.

| Vulnerability | Safe | Vulnerable | Total |
| --- | --- | --- | --- |
| Reentrancy | 90 | 95 | 185 |
| Suicide Attacks | 89 | 220 | 309 |
| Integer Overflow/Underflow | 96 | 63 | 159 |
| **Total** | **275** | **378** | **653** |

*3.1.2 Data Cleanup.* We automatically remove non-ASCII artifacts, correct missing or inconsistent pragma declarations using the metadata that we record for each contract. Additionally, we isolate individual sub-contracts from composite JSON files to ensure successful compilation. After this cleanup, we obtain 653 contracts that compile and run correctly across all studied analyzers.

*3.1.3 Labeling and Validation.* Following the principle "vulnerable does not imply exploited" [46], our labeling distinguishes between theoretical and exploited vulnerabilities. To achieve that, we first manually inspect the source code of each contract to identify vulnerability patterns. For reentrancy, we check whether functions use the method call() to transfer funds to external contracts before updating state variables. This is because such pattern usually leads to recursive calls, which is a common pattern in reentrancy attacks. We then review the transaction history of the contract to detect on-chain evidence of exploitation. For suicide attacks, we check for unauthorized calls to selfdestruct(), and review transaction logs for on-chain evidence of contract destruction. For arithmetic errors, we identify arithmetic operations that may exceed the bounds of a variable's type without using secure libraries (e.g., SafeMath [43]).

This process yields ScBench, a dataset of 275 safe and 378 vulnerable contracts. Table 1 shows their distribution, where each contract contains at least one of the studied vulnerability types. ScBench reflects realistic contract states and vulnerability patterns, enabling precise measurement of false-positives and false-negatives.

### 3.2 Security Analyzer Selection

Our selection of security analyzers builds upon the comprehensive, peer-reviewed catalogs compiled by Ivanov et al. [32] and Zhu et al. [64], the most recent and extensive surveys published in 2023 and 2024, respectively, rather than a popularity-based selection. Together, these surveys enumerate 133 and 178 distinct papers and tools, providing a methodologically diverse and empirically grounded foundation. From this broad inventory, we apply the following inclusion criteria:

- *Open availability*: the analyzer must be publicly accessible and runnable through a Command Line Interface (CLI) to enable large-scale experimentation.
- *Version compatibility*: the tool must support Solidity compiler versions 0.4.x–0.8.x to ensure coverage of commonly deployed contracts.
- *Multi-class coverage*: the analyzer must detect at least two of the three vulnerability classes in SCBENCH.
- *Practical adoption*: the tool should demonstrate active use in research or auditing practice, reflected through citations, community engagement, or repository activity (e.g., GitHub > 1000 stars and > 300 forks).

To ensure methodological consistency and reproducible evaluation across different analysis paradigms, we exclude the following categories of tools:

- Proprietary or subscription-based analyzers (e.g., Zeus [34]), as their closed-source nature prevents transparent evaluation and replication of results.
- Tools targeting outdated Solidity versions (e.g., Sailfish [8]), which are incompatible with current compilers and would introduce operational bias.
- Tools requiring user defined specifications (e.g., Echidna [29]), since their performance depends heavily on manual input, undermining objective, automated comparison.
- Analyzers detecting only a single vulnerability type (e.g., eThor [51], SaferSC [56]), as they do not support a comprehensive evaluation across our selected vulnerability classes.

We further exclude LLM-based analysis techniques, because they generally do not support the automated, reproducible evaluation required by our study. Most current LLM-based tools are published as research prototypes tailored to specific datasets, making them unsuitable for integration into a standardized benchmarking pipeline. For example, we attempt to evaluate BlockScan [62], but its implementation is tightly coupled to its transactional dataset and we could not adapt it to our benchmark. We have reached out to the authors for guidance, but have received no response. Similarly, we could not obtain a runnable version of BlockGPT [25] due to the lack of a reusable publicly available tool. These practical barriers prevent the consistent and fair comparison necessary for our evaluation. We do not exclude ML-based tools categorically; for example, we include Dlva [3] because it meets our criteria, but exclude others (e.g., SaferSC [56]) that lack broad vulnerability coverage.

Applying our inclusion and exclusion criteria yields:

- CONFUZZIUS [57]: a hybrid fuzzing tool for Ethereum smart contracts that combines data-dependency analysis with traditional fuzz testing. It uses taint tracking and lightweight symbolic execution to guide input generation towards potentially vulnerable paths, improving detection of complex issues such as reentrancy and arithmetic bugs. Confuzzius balances fuzzing scalability with deeper bug coverage by focusing on dependency-aware input selection.
- DLVA [3]: a deep learning-based analyzer that operates on Ethereum Virtual Machine (EVM) bytecode using a Graph Neural Network (GNN) to extract structural features. Since Dlva operates on EVM bytecode, it does not need access to the original Solidity source code of a contract. However, Dlva is trained on contracts labelled by SLITHER [24], inheriting biases from those labels.
- MYTHRIL (v0.24.7) [39]: a widely adopted symbolic execution engine for analyzing EVM bytecode. Mythril systematically explores execution paths with symbolic transaction parameters, builds control-flow graphs, and uses the Z3 SMT solver [16] to identify vulnerabilities such as reentrancy, suicid attacks, and arithmetic bugs. It produces concrete exploit traces, offering actionable outputs alongside vulnerability reports.
- OSIRIS [58]: a symbolic execution-based tool that detects integer-related vulnerabilities (e.g., overflows and underflows). Osiris targets arithmetic-specific invariants through focused SMT queries, offering precise detection even in complex control flows (e.g., loop-heavy contracts).
- OYENTE [37]: one of the first symbolic execution frameworks for Ethereum smart contracts. Oyente detects issues such as reentrancy, transaction-ordering dependence, and unchecked call returns. It uses path feasibility checks to reduce false positives and can generate concrete test inputs. Oyente's modular design laid the foundation for subsequent symbolic execution tools and continues to be influential in both academic research and security audits.
- SLITHER (v0.10.4) [24]: a fast static analyzer that operates on Solidity source code. Slither constructs SlitherIR, its own Intermediate Representation (IR), and then applies more than 70 vulnerability and quality checks (e.g., reentrancy, integer bugs, improperly enforced access controls). Its lightweight performance enables analyzing contracts in 2–3 seconds, making it practical for large-scale detection and integration into Continuous Integration (CI) pipelines.

Table 2: Empirical results for reentrancy.

| Tool | P | R | F1 | FNR | FPR | Time (s) |
|---|---|---|---|---|---|---|
| CONFUZZIUS | **94.8%** | **94.6%** | **94.6%** | 2.1% | 8.9% | 565.12 |
| DLVA | 30.2% | 43.8% | 31.2% | 97.9% | 12.2% | 2.51 |
| MYTHRIL | 64.6% | 55.7% | 49.0% | 80.0% | **6.7%** | 723.66 |
| OSIRIS | 86.5% | 84.3% | 84.0% | 3.2% | 28.9% | 117.25 |
| OYENTE | 87.3% | 85.4% | 85.2% | 3.2% | 26.7% | 7.91 |
| SLITHER | 91.9% | 90.8% | 90.7% | **1.1%** | 17.8% | **1.33** |

### 3.3 Experimental Setup and Metrics

To collect the results for our analysis, we run each tool on SCBENCH, with a timeout of 15 minutes per contract. For every run, we record the reported vulnerability type and runtime. We then compare the output against the ground truth labels to assess their performance in terms of precision (P), recall (R), F1-score (F1), False Negative Rate (FNR), False Positive Rate (FPR), and average analysis time per contract. These metrics provide insights into each tool's accuracy, ability to identify vulnerabilities, and computational efficiency.

### 3.4 Results

*3.4.1 Reentrancy.* Table 2 shows that CONFUZZIUS achieves the highest overall accuracy (F1 = 94.6%), leveraging its hybrid fuzzing and taint-tracking design. Slither follows (F1 = 90.7%) with near-instantaneous results (1.33 seconds per contract) but a higher FPR (17.8%). Symbolic execution tools (Oyente and Osiris) are moderately precise (F1 = 85.2% and 84%, respectively) but are more prone to over-flagging (FPR = 26.7% and 28.9%, respectively) than Slither. Mythril performs the weakest (F1 = 49%, FNR = 80%), mislabeling most vulnerable contracts. Dlva shows similar weaknesses, achieving modest precision (F1 = 31.2%), and missing nearly all true positives (FNR = 97.9%) due to its conservative classification thresholds that reduce FPR (12.2%) at the cost of recall.

Table 3: Empirical results for suicide attacks.

| Tool | P | R | F1 | FNR | FPR | Time (s) |
|---|---|---|---|---|---|---|
| Confuzzius | **78.1%** | 51.1% | 50.3% | 66.8% | **4.5%** | 481.01 |
| Dlva | 57.0% | 40.1% | 40.4% | 70.9% | 32.6% | 2.20 |
| Mythril | 77.5% | 49.2% | 47.8% | 69.5% | **4.5%** | 491.71 |
| Slither | 70.0% | **57.6%** | **59.4%** | **49.1%** | 25.8% | **1.71** |

Table 4: Empirical results for integer arithmetic errors.

| Tool | P | R | F1 | FNR | FPR | Time (s) |
|---|---|---|---|---|---|---|
| Confuzzius | 47.5% | 55.3% | 47.9% | **12.7%** | 14.6% | 304.38 |
| Mythril | 77.8% | 64.8% | 54.7% | 88.9% | **0.0%** | 664.07 |
| Osiris | **84.1%** | **83.0%** | **83.2%** | **12.7%** | 19.8% | 240.04 |
| Oyente | 75.7% | 75.5% | 75.6% | 28.6% | 21.9% | **17.61** |

*3.4.2 Suicidal Contract Termination.* Table 3 shows that all tools struggle with this vulnerability type. Slither leads (F1 = 59.4%) but its 25.8% FPR inflates the cost of manual triage that developers have to do. Mythril and Confuzzius miss roughly two-thirds of true cases, showing that their heuristics poorly model authorization logic. Dlva achieves a moderate precision (P = 57.0%) and low recall (R = 40.1%), missing a large fraction of true positives (FNR = 70.9%) while also generating substantial false positives (FPR = 32.6%).

With respect to analysis time, Slither remains the fastest at an average speed of 1.71 seconds per contract, whereas Confuzzius and Mythril require >480 seconds, highlighting their computational intensity. Dlva is orders of magnitude faster than Confuzzius and Mythril, and a close second to Slither. However, Dlva achieves that at a clear cost to detection quality.

*3.4.3 Integer Arithmetic Errors.* Table 4 shows that Osiris, specialized for arithmetic invariants, attains the best F1 (83.2%) and low FNR (12.7%), followed by Oyente (F1 = 75.6%). Confuzzius and Mythril trail far behind (F1 = 54.7% and 47.9%, respectively), highlighting their limited support for numeric reasoning.

> No single analyzer dominates across all vulnerability types. Confuzzius excels at detecting reentrancy, while Osiris performs best on integer arithmetic errors. Slither and Oyente trade their accuracy for faster scans. Dlva broadens coverage to contracts without source code, but still lag behind source-aware analyzers. Suicide contract termination is the most challenging to detect (F1 <60%).

## 3.5 Discussion

ScBench exposes a clear accuracy–efficiency frontier: precise detection often comes with prohibitive runtime, while faster tools over-flag benign contracts. These findings reveal why developers, even when tools exist, hesitate to integrate them into their workflows. To address these challenges, we propose the following targeted recommendations, each grounded in our findings:

(1) *Integrate complementary approaches for broader coverage.* Confuzzius has the highest detection rates for reentrancy, reflecting the effectiveness of combining fuzzing with symbolic execution. Extending this hybrid analysis approach to symbolic execution tools (e.g., Mythril) may reduce false negatives by improving path exploration in complex contracts.
(2) *Optimize symbolic execution.* Mythril has high FNR, and runs the slowest due to its exhaustive symbolic execution. Techniques such as bounded exploration [11], feasibility checks similar to Oyente, or SMT query optimization [15] may improve both efficiency and vulnerability coverage.
(3) *Expand specialized vulnerability patterns.* Thanks to its specialized arithmetic predicates, Osiris has the best performance in detecting integer arithmetic errors. Incorporating similar tailored vulnerability patterns into other tools (e.g., Slither and Confuzzius) may enhance their detection capabilities for integer arithmetic errors and suicide attacks.
(4) *Reduce false positives via cross-validation.* High FPR for Osiris and Oyente undermine trust in their outputs. Integrating static analysis results from other tools (e.g., Slither) as a cross-validation step would help filter implausible vulnerability reports while maintaining acceptable analysis times.
(5) *Invest in scalability.* Confuzzius and Mythril have substantially longer analysis times compared to other tools. Parallelizing symbolic execution and fuzzing workloads, as well as incorporating static pre-analyses to prioritize high-risk paths, would help overcome this limited scalability. Techniques such as bounded symbolic execution [11], feasibility checks similar to Oyente, or SMT query optimization [15] can also be used to improve both efficiency and vulnerability coverage. Improving scalability is key to make deep analyses practical within developers' expected runtime budgets (≤ 10 minutes).
(6) *Develop ensemble-based analyzers.* No single tool consistently outperforms others across all vulnerability types and performance metrics. Developing ensemble frameworks that combine outputs from multiple analyzers (e.g., in a game-theoretic fashion [30]) would leverage their strengths, provide broader coverage, and reduce the impact of individual weaknesses.
(7) *Develop and release LLM-based detectors.* Researchers should prioritize the public release of LLM-based vulnerability detectors that meet reproducibility criteria, including benchmark-ready artifacts and evaluation scripts to enable fair, automated comparison.
(8) *Expand benchmarking to emerging threats.* Our evaluation excludes flash-loan and oracle-manipulation vulnerabilities due to limited analyzer support and low developer familiarity. As tools mature, future benchmarks should incorporate curated datasets for these exploit types to assess detection capabilities against modern attack vectors.

## 4 Surveying Smart Contract Developers

The benchmark results in Section 3 reveal significant variability in the accuracy, efficiency, and reliability of existing analyzers. To understand how these technical limitations influence developer trust and adoption, we complement our quantitative evaluation with a survey of professional smart contract developers and auditors. This section reports the qualitative and quantitative findings from that survey, which forms the second strand of our convergent parallel mixed-methods design [54].

### 4.1 Survey Design

*4.1.1 Questionnaire.* The survey comprises 25 questions that collect demographic and contextual information, tool-usage patterns, perceptions of accuracy and trust, and preferences for explanation formats. The questions are a mix of multiple-choice, Likert-scale, and open-ended formats to balance quantitative measurement with qualitative depth. Table 5 lists all survey questions. The complete questionnaire and anonymized responses are publicly available [1].

*4.1.2 Recruitment.* We recruited participants through Prolific [47], pre-screening for involvement in smart-contract development, auditing, or blockchain security. Our inclusion criteria is:

- primary language is English,
- Prolific approval rate between 95–100%,
- current work involves programming languages, and
- holding an organizational role in data analysis, software engineering, or information security.

Over a three-week data collection period on Prolific, we received 347 completed survey responses. Applying established quality-control practices for the platform [44], we filtered this set to 150 high-quality submissions from participants with genuine experience in smart-contract analysis.

To verify participant expertise, we first use Prolific's pre-screening filters to recruit individuals who self-identify as professionally involved in smart contract development, auditing, or blockchain security. Additionally, we have competency-check questions within the survey itself. In particular, Q6 tests the ability to recognize both classical and less common vulnerability types, while Q7 probes for hands-on experience with specific analysis tools. We remove responses that show inconsistent knowledge, such as claiming proficiency while failing to identify fundamental vulnerabilities, or exhibit bot-like patterns. This approach evaluates applied understanding rather than relying solely on self-reported experience. The high inter-rater reliability in our subsequent qualitative coding (Cohen's $\kappa$ = 0.755–0.866) further confirms that retained participants provided coherent, expert-level feedback. Each verified participant received £5 as a fair monetary compensation, according to the average compensation on Prolific, for their time.

*4.1.3 Ethical Considerations.* Our study was approved by the Institutional Review Board (IRB) at NYU Abu Dhabi (Study ID: HRPP-2025-3). The study is considered to pose minimal risk to participants. Before answering the questions, we obtain informed consent from each participant. We inform participants with the purpose of our survey, and Prolific allows them to quit the survey at any time. We ensure that the collected data is anonymized. We also pilot the survey to ensure that it can be completed within a reasonable amount of time (25–30 minutes).

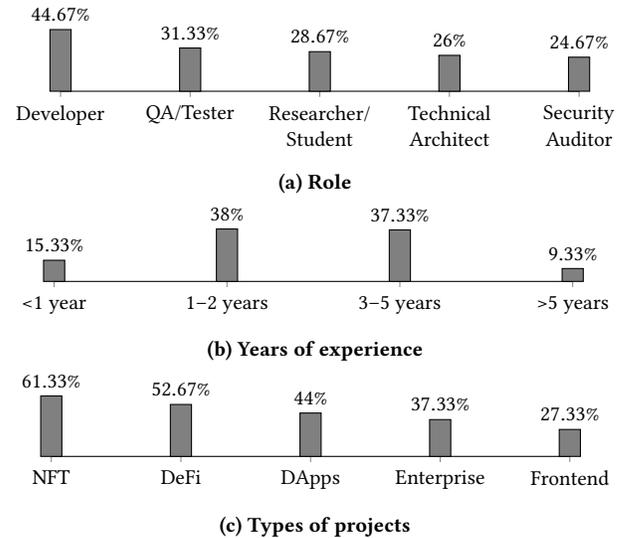

Figure 1: Background of survey participants.

### 4.2 Research Questions

The main research question that motivates our work is: *are existing security analyzers not sufficiently effective, or are users not using them effectively (if at all)?* We break down this main research question into several sub-questions:

- **RQ1**: What are the most commonly recognized security vulnerabilities among users?
- **RQ2**: What are the most commonly used security analyzers?
- **RQ3**: What is an acceptable runtime performance of a security analyzer?
- **RQ4**: What are the factors that affect trust/confidence in security analyzers?
- **RQ5**: Why do users ignore reported vulnerabilities by security analyzers?
- **RQ6**: What types of explanations for vulnerabilities flagged by security analyzers do users find most effective in enhancing trust and facilitating remediation?

### 4.3 Findings

To analyze the quantitative survey responses, we employ descriptive statistics. For qualitative insights, two authors have conducted a Modified-Delphi card sorting [45] to systematically categorize open-ended responses.

*4.3.1 Participant Background.* Figure 1 shows that participants represent diverse roles: developers (44.7%), testers (31.3%), researchers/students (28.7%), technical architects (26%), and auditors (24.7%). Most participants (75.33%) have 1–5 years of experience, and a majority works on NFT (61.33%) and DeFi (52.67%) projects. Finally, most participants (96%) perceive smart contract security as important.

Table 5: Survey questions grouped by section.

| Section | ID | Question |
|---|---|---|
| Participant Background | Q1 | What is your role in the smart contract development process? |
| | Q2 | How many years of experience do you have with smart contract development? |
| | Q3 | Which programming languages do you primarily use for smart contract development? |
| | Q4 | What type of smart contract projects are you primarily involved in? |
| | Q5 | How important is security and safety in your smart contract development process? |
| | Q6 | Which of the following smart contract security vulnerabilities are you familiar with? |
| Usage of Security Analyzers | Q7 | Which security analyzers have you used during smart contract development? |
| | Q8 | How frequently do you use security analyzers when developing smart contracts? |
| | Q9 | At which stages of development do you typically use security analyzers? |
| | Q10 | What are your main reasons for using security analyzers? |
| | Q11 | Which type of interface do you prefer for using a security analyzer? |
| | Q12 | What type of input do you typically analyze with a security analyzer? |
| | Q13 | What is your preferred pricing model for a security analyzer? |
| | Q14 | What is the longest amount of time you would typically allow a security analyzer to run before expecting results? |
| | Q15 | On average, how much time do you spend verifying whether a reported vulnerability is a true positive? |
| Confidence in Security Analyzer Outputs | Q16 | How confident are you in the accuracy of vulnerabilities reported by security analyzers? |
| | Q17 | Which of the following factors increase your confidence in a security analyzer's results? |
| | Q18 | Which of the following factors reduce your confidence in a security analyzer's results? |
| | Q19 | Have you ever ignored a vulnerability reported by a security analyzer? |
| | Q20 | Can you briefly explain a situation where you chose to ignore a reported vulnerability? |
| Impact of Explanation on Confidence | Q21 | What kinds of explanation do you find most helpful from a security analyzer when it reports a vulnerability? |
| | Q22 | To what extent does the quality of the explanation affect your confidence in the analyzer's results? |
| | Q23 | If a security analyzer reports a vulnerability and provides a detailed explanation with a code snippet or suggested fix, how confident are you that the vulnerability is a true positive? |
| | Q24 | If a security analyzer reports a vulnerability but does not include an explanation, how confident are you that it is a true positive? |
| | Q25 | Do you have any suggestions for improving the explanation formats used by security analyzers? |

4.3.2 Commonly Recognized Vulnerabilities (**RQ1**). Figure 2 shows that participants are most familiar with classical vulnerabilities such as reentrancy (62.7%) and integer arithmetic errors (54.7%), while awareness of emerging threats such as flash-loan attacks or oracle-manipulation attacks remains below 40%. This uneven familiarity suggests that education and tool documentation continue to emphasize established vulnerabilities, leaving new exploit classes under-represented.

> **RQ1**: Users are most familiar with classical vulnerabilities such as reentrancy, while newer threats such as flash loan and oracle manipulation are less recognized, highlighting limited awareness of emerging attack types.

4.3.3 Usage Patterns (**RQ2**). Since most participants value smart contract security, more than half (54.66%, Q8) use security analyzers for more than 75% of their work. The uses (Q9) vary from during audits (78%) to coding (75.33%) and testing (74%). The primary reasons (Q10) for using security analyzers include identifying vulnerabilities (89.33%), ensuring code quality (82%), learning about potential security issues (68%), and complying with organizational or regulatory policies (53.33%). Participants have no clear preference of tool interfaces (Q11). While 34% prefer CLI, 28% prefer Integrated Development Environment (IDE) plugins, and 21.33% would rather use a web-based interface. With respect to tool cost, a majority of participants prefer to use freemium pricing models (52%, Q13).

Figure 3 shows that SLITHER is the most commonly used analyzer (77.33%), followed by MYTHRIL (51.3%). Their popularity reflects practical considerations. SLITHER's active maintenance, fast analysis, and broad vulnerability coverage (as we show in Section 3) make it a default choice for many developers. Nevertheless, MYTHRIL remains attractive for its symbolic-execution capabilities despite slower runtimes. In contrast, analyzers that demand extensive user-defined inputs (e.g., Echidna) or that are no longer actively maintained (e.g., SmartCheck) exhibit notably lower adoption.

Although the literature contains many Machine Learning (ML)-based vulnerability detectors [64], few authors release production-ready tools or datasets, and none of our survey participants reported using ML-based tools. This absence highlights a gap between research and practice likely caused by limited tool availability, integration and deployment challenges, and concerns about explainability and trust in ML-driven findings.

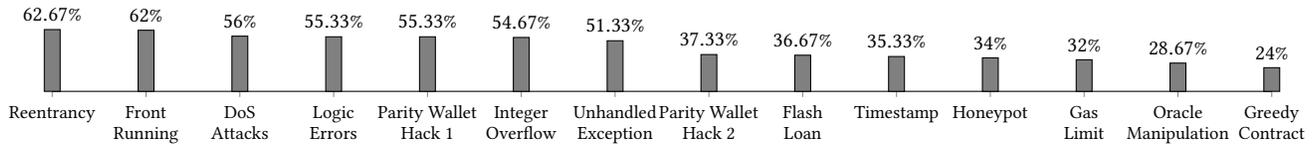

Figure 2: Familiarity with smart contract security vulnerabilities.

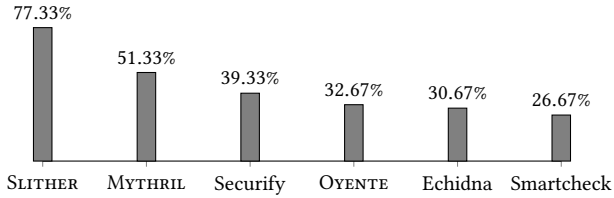

Figure 3: Most commonly used security analyzers.

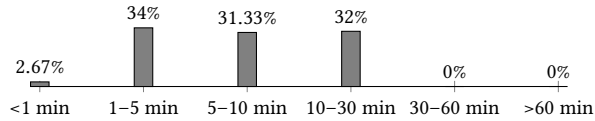

Figure 4: Acceptable running time for security analyzers.

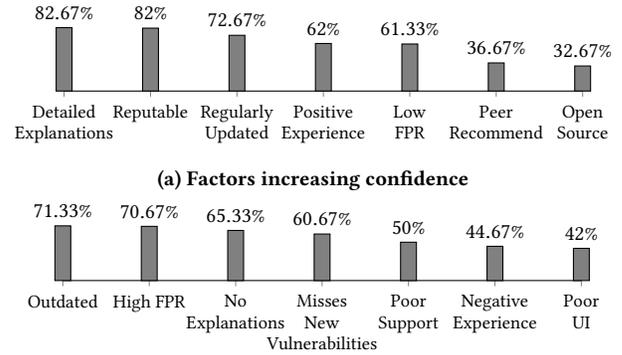

Figure 5: Factors affecting confidence in security analyzers.

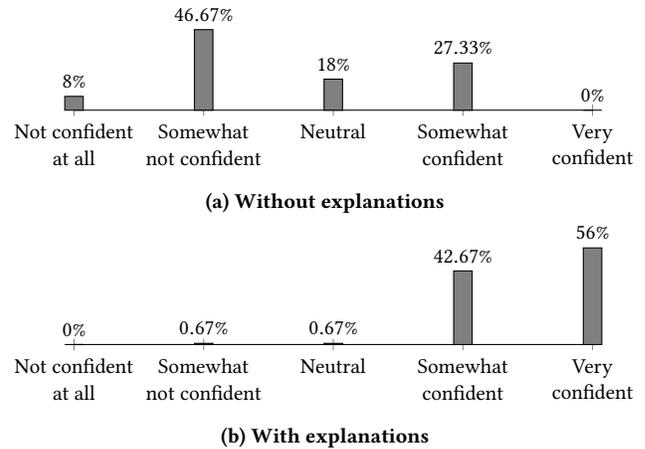

Figure 6: Confidence in security analyzers.

> **RQ2**: Slither and Mythril are the most widely used analyzers, reflecting developers' preference for tools that are actively maintained, fast, and easy to integrate into their workflows.

*4.3.4 Acceptable Runtime Performance (RQ3).* Figure 4 shows that more than two-thirds of participants expect analyzer results within 10 minutes (34% prefer 1–5 minutes, 31% prefer 5–10 minutes), and none accept runtimes beyond 30 minutes. These expectations mirror the practical runtime limits that we observe in Section 3, underscoring the importance of balancing accuracy with timely feedback for integration into development workflows.

> **RQ3**: 68% of participants expect analyzer results within 10 minutes, mirroring the time limits that we observe in our empirical evaluation.

*4.3.5 Factors Affecting Confidence in Analyzers (RQ4).* Most participants (73%) report partial trust in analyzer outputs (Q16), and often review serious findings manually. Figure 5 presents the main factors that participants find helpful to improve their confidence in tool results: detailed explanations (82.67%) and analyzer reputation (82%). Conversely, confidence is undermined by outdated tools (71.33%), high FPR (70.67%), and lack of explanations (65.33%). The quality of explanations influences confidence for 58.67% of participants. These findings echo the technical shortcomings that we identify in Section 3.

Figure 6 further shows that trust in analyzer results increases sharply when they provide explanations or supporting evidence: 98.67% of participants report higher confidence when explanations accompany alerts, and half indicate at least a two-level improvement on a five-point scale.

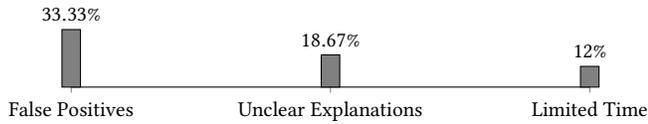

Figure 7: Main reasons to ignore analyzer warnings.

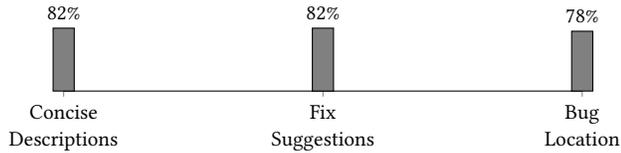

Figure 8: Preferred explanation formats.

> **RQ4**: The dominant trust barriers in analyzer confidence are high false-positive rates (71%), outdated tool support (71%), and lack of explanations (65%). An overwhelming majority of users (98%) feel more confident when tools provide clear, detailed explanations of their results.

*4.3.6 Reasons To Ignore Reported Vulnerabilities (RQ5).* To investigate why users disregard vulnerabilities flagged by smart contract security analyzers, our survey has closed-ended (Q19) and open-ended (Q20) questions. Approximately one-third of participants (34.67%) never ignore warnings. However, Figure 7 shows that users ignore warnings primarily due to excessive false positives (33.33%) or unclear explanations (18.67%). A smaller portion (12%) cite limited time as a reason for dismissal.

To gain deeper insights, we conduct a qualitative analysis of the open-ended responses to Q20, coding prevalent themes to identify common rationales. The inter-rater reliability for coding these responses is high (Cohen's $\kappa$ = 0.866), exceeding the 0.81 threshold for almost perfect agreement [36], ensuring the robustness of our qualitative analysis. Across the responses to Q20, the most frequent themes are "manual verification" (14.67%) and "false positives" (14%). Among mixed responses, the top two themes are also "false positives" (29.17%) and "manual verification" (27.78%). Since those two themes are indicators of mistrust in tool accuracy, our qualitative analysis reinforces that trust degradation, not indifference, drives users to ignore reported vulnerabilities.

> **RQ5**: Users ignore security warnings primarily due to excessive false positives (33.3%) and unclear explanations (18.7%). While a third of participants never ignore alerts, for most, dismissing them is a pragmatic response to tool inaccuracy and the high effort required to validate findings.

*4.3.7 Most Useful Types of Explanations (RQ6).* To understand user preferences for vulnerability explanation formats, we analyze responses to survey questions Q21 and Q25. Figure 8 shows that participants equally prefer concise vulnerability descriptions and fix suggestions (82% each), closely followed by a preference for precise bug locations (78%).

To identify specific preferences, we further conduct a qualitative analysis of the responses to Q25. The inter-rater reliability for coding these responses is relatively high (Cohen's $\kappa$ = 0.755), which is close to the 0.81 threshold that typically indicates substantial agreement between independent annotators [36]. Standalone preferences are not that common in the responses, and include features such as "structured explanations" (8%), "contextualized explanations" (6.67%), "actionable remediation steps" (6%), and "visual aids" (3.33%). The majority of responses (58.67%) exhibit a *mixed rationale*, indicating a desire for a combination of explanation features. Within these mixed responses, the most frequent are "actionable remediation steps" (20.15%) and "contextualized explanations" (17.16%), followed by "real-world examples and documentation" (17.16%), "visual aids" (13.81%), and "structured explanations" (11.94%). These findings indicate that users value actionable and clear explanations that are not only concise but also practical and directly applicable to resolving the identified vulnerabilities.

> **RQ6**: Developers prioritize actionable and efficient vulnerability explanations. Most users prefer brief descriptions with fix suggestions (82%) and precise bug locations (78%). When they require more details, participants often prefer actionable remediation steps (20.2%), contextualized explanations (17.2%), and real-world examples (17.2%).

## 4.4 Implications for Tool Design

Integrating user feedback with our benchmarking results yields several actionable recommendations for designing security analyzers for smart contracts. These recommendations directly reflect user needs and expectations, supporting more effective, trustworthy, and widely adopted tools.

(1) *Balance accuracy with feedback speed.* More than two thirds of participants expect analysis results within 10 minutes, and none accept tools to take 30 minutes or more to produce their results. To integrate into iterative development workflows without disrupting productivity, analyzers should aim for that 10-minute window.

(2) *Prioritize clarity over verbosity.* The quality of reported explanations affect the confidence of more than half of participants in the produced analyzer results. Additionally, more than 80% of participants prefer brief descriptions and actionable fix suggestions. Therefore, analyzers should focus on providing succinct, verifiable explanations, and concrete remediation steps for flagged vulnerabilities.

(3) *Improve precision and filtering.* The main reason that participants ignore warnings is high FPR (33.33%). Since high FPR directly translates into wasted triage effort and reduced trust, security analyzers must focus on producing precise results.

(4) *Support flexible interfaces.* Participants prefer analyzers available as CLI tools, IDE plug-ins, or web dashboards, reflecting diverse workflows. To maximize their adoption, analyzers should support flexible deployment options that accommodate those diverse workflows.

(5) *Reconcile conciseness with contextual depth through mode-based explanations.* Our findings show that developers prefer concise descriptions (82%) but also request detailed context such as actionable remediation steps (20.2%). Rather than a contradiction, these preferences reflect different workflow phases. During triage, developers need brief one-line descriptions with bug locations for quick go/no-go decisions. During remediation, they require expandable, detailed fix suggestions when a warning is deemed credible. An ideal analyzer should support both modes: a *summary* mode providing brief descriptions for scanning many warnings, and a *detail* mode with code snippets and remediation patterns for confirmed issues, analogous to compiler error messages with optional `-explain` flags.

These practitioner insights, when combined with the quantitative benchmarking, reveal how measurable tool deficiencies manifest as usability and trust barriers in practice. Together, they provide an evidence-based foundation for designing next-generation analyzers that developers are both willing and able to use.

## 5 Threats to Validity

### 5.1 Internal Validity

A primary threat arises from the accuracy of ground truth labels for vulnerabilities in our curated dataset. Inaccuracies in these labels might bias our valuation metrics such as precision and recall, potentially skewing our assessment of tool performance. To mitigate this, we conducted a manual code analysis to identify vulnerability patterns, reviewed transaction histories to detect evidence of exploitation, and cross-validated labels against reported incidents from security audits and academic literature [3, 14, 27, 31] and ensured that the dataset includes both exploited and unexploited contracts, adhering to the principle: *vulnerability does not imply exploitation* [46].

Another threat is the 15-minutes timeout set for tool evaluations, which might truncate analysis for slower tools (e.g., Mythril and Confuzzius), affecting performance metrics. However, this limit is 50% more than the 10-minute window within which developers expect to receive analysis results. Therefore, we see it is a reasonable balance between completeness and practical constraints.

Finally, self-reported data from our survey may be subject to social desirability bias, potentially inflating reported expertise or tool usage. To address this issue, we ensured anonymity through Prolific [47] and targeted experienced practitioners to encourage candid responses.

### 5.2 External Validity

Our dataset of Ethereum contracts may not fully represent contracts on other blockchains (e.g., Binance Smart Chain [7]), potentially limiting the applicability of our results. To address this threat, we ensured that our dataset included diverse vulnerability types and contract states that reflect real-world scenarios.

Similarly, our survey's focus on Ethereum developers may restrict broader applicability to other blockchain ecosystems. To mitigate this threat, we recruited a diverse sample of practitioners with different roles and experience levels, representing a broad cross-section of the Ethereum community. Our study focuses on Ethereum, because it hosts the largest, most mature smart-contract ecosystem, providing abundant real-world incidents, diverse deployed bytecode, and well-established analysis tooling. These conditions are necessary for a thorough, reproducible evaluation. Our findings are directly applicable to EVM-compatible chains (e.g., Binance Smart Chain [7], Polygon [35]), because they share the same bytecode, compiler toolchain, and most vulnerability classes. Extending the study to non-EVM platforms (e.g., Solana [61]) requires retooling the pipeline with different compilers, language semantics, vulnerability taxonomies, and chain-specific labeling rules, which we leave for future work.

Furthermore, our findings for the six tools evaluated may not extend to other tools or future versions. To counter this threat, we selected widely-used, actively maintained tools employing diverse analysis techniques.

### 5.3 Construct Validity

A key threat is that our definitions of vulnerabilities may differ across practitioner interpretations, potentially misaligning our metrics with real-world perceptions. To mitigate this threat, we adopted standardized definitions from established literature [4, 63]. We also focused on well-documented issues validated by security audits.

Another concern is that user trust, measured through self-reported confidence and qualitative responses, may not fully reflect actual behavior. To address this issue, we combine quantitative scales with qualitative open-ended questions.

While metrics such as precision, recall, and F1-score evaluate technical performance, they may not fully capture usability or real-world effectiveness. To address this issue, we incorporated multiple metrics, including false negative rate, false positive rate, and analysis time. We then complemented these metrics with survey insights on trust and usability to provide a comprehensive assessment.

## 6 Related Work

### 6.1 Smart Contract Vulnerabilities

Research on smart contract vulnerabilities has established a foundation for understanding their nature and impact. Atzei et al. [4] provide a seminal taxonomy of common vulnerabilities, including reentrancy, integer overflow/underflow, and unchecked external calls, which continue to threaten contract integrity. Zhou et al. [63] builds on this work by applying it to smart contract security, identifying over 20 vulnerability types and emphasizing the need for automated detection mechanisms. Real-world incidents (e.g., the DAO hack in 2016 [52] and the suicide attack in 2017 that froze US$150 million [50]) illustrate the severe consequences of these flaws. Our study leverages a dataset of 653 real-world contracts exhibiting reentrancy, suicide, and integer overflow/underflow vulnerabilities, reflecting these well-documented issues to comprehensively evaluate security analyzers.

### 6.2 Empirical Evaluation of Smart Contract Security Analyzers

Several large-scale empirical assessments have established benchmarks for tool performance. Durieux et al. [20] have conducted an extensive evaluation of nine tools on 47,587 real-world unlabelled contracts, and 69 annotated vulnerable contracts, establishing the reproducible SmartBugs framework. While this work highlights detection overlaps and inconsistencies across tools, the evaluation is limited to positive (i.e., vulnerable) samples, which inherently restricts the ability to fully assess tool performance metrics such as False Positive Rate (FPR) and False Negative Rate (FNR) by excluding a balanced set of non-vulnerable (i.e., negative) contracts.

To address tool limitations through systematic testing, Ghaleb and Pattabiraman [26] propose SolidiFI, a mutation testing framework. This approach injects 9,369 targeted vulnerabilities into 50 contracts to measure the false negatives and positives of six static analyzers. Although the results revealed several tool limitations, they are limited to the use of synthetic, injected bugs, which may not capture the full complexity and diversity of real-world vulnerabilities and developer code.

Other studies have focused on the discrepancy between flagged vulnerabilities and actual exploitation. Perez and Livshits [46] examine this gap using Datalog queries across 23,327 vulnerable contracts and 20 million Ethereum transaction traces, revealing that only 1.98% of the flagged contracts are actually exploited. This work offers crucial insight into why exploitation rates are low (e.g., fund concentration). However, it relies on aggregated prior vulnerability data without re-executing the tools or including negative samples, thereby limiting the calculation of standard performance metrics such as FPR and FNR.

More recently, Chaliasos et al. [13] present an evaluation of five prominent analysis tools against 127 documented high-impact DeFi attacks, totaling over US$2.3 billion in losses, combined with a survey of 49 practitioners. The authors found that the tools could have prevented only 8% of these attacks, primarily reentrancy-based issues, highlighting significant industry gaps in detecting logic vulnerabilities. However, their evaluation centered on highly specialized exploits (e.g., flash loan-related) where 75% of them fall outside the current scope of their benchmarked tools. Additionally, the reported evaluation does not assess false positives, false negatives, or efficiency metrics using a comprehensive and balanced dataset.

In contrast to prior work, our study provides a comprehensive performance and efficiency benchmark of six widely-used tools on a tailored dataset of 653 manually labeled real-world contracts. We precisely measure precision, recall, FPR, FNR, and efficiency across three critical, high-monetary-loss vulnerability types: reentrancy, suicide, and integer overflow/underflow. Our measurements reveal high FPRs (e.g., 17.8% for Slither on reentrancy, 26.7% for Oyente) that impose a significant, unsustainable verification burden on developers. This finding is directly corroborated by our large-scale survey of 150 practitioners, where 33.33% cite false positives as a reason for ignoring security warnings. The deliberate focus of our study on three vulnerability classes, paired with a custom-labeled dataset enables a rigorous, consistent, and integrated analysis of both tool performance and the resulting human-centric challenges.

### 6.3 User Trust and Usability in Smart Contract Security

While the technical capability of security tools is paramount, user trust and usability are critical human factors that directly influence the practical adoption and effectiveness of these tools in development workflows. Chaliasos et al. [13] have explored developer experiences by surveying 49 practitioners on general tool usage and workflow challenges. However, the study does not focus on how quantitative performance metrics (e.g., FPR, FNR, analysis time) impact user trust in the reported vulnerabilities. While numerous other surveys and cataloging efforts exist for smart contract tools and techniques [32], they seldom address human-centric adoption or trust challenges.

In the broader software engineering domain, Noller et al. [40] surveyed over 100 practitioners to identify factors that increase trust in automatically generated patches for general-purpose programming languages, with an emphasis on preferences for interaction and supporting evidence. Whether those findings are transferable to the high-stakes, domain-specific context of smart contracts—where missed vulnerabilities can have catastrophic consequences—remains unexplored. Our work fills this gap by specifically investigating developer trust in smart contract security tools. We move beyond general tool usage to quantify how high false alarm rates, low detection rates, the lack of clear vulnerability explanations, and poor tool integration erode developer confidence, often leading to warning fatigue and the dismissal of critical alerts. This comprehensive approach, integrating technical performance data with deep qualitative insights into developer trust, is a distinguishing factor of our contribution.

## 7 Conclusion

This paper presents a large-scale, practitioner-grounded assessment of smart-contract security analyzers that integrates quantitative benchmarking and qualitative developer insights. Our evaluation of 653 hand-validated contracts (ScBench) reveals substantial variability in analyzer accuracy, efficiency, and usability. F1 scores range from 31.2% to 94.6%, false-positive rates reach up to 32.6%, and some analysis times exceed 700 seconds per contract. No single tool dominates across all vulnerability classes. For example, Confuzzius excels on reentrancy, while Slither achieves the lowest false-negative rate, yet both underperform on suicidal contract termination and arithmetic errors.

Our survey of 150 professional developers complements these empirical findings by exposing the human factors behind low adoption. Users cite excessive false positives, unclear explanations, and slow feedback as the main barriers to trust. Collectively, these insights demonstrate that the technical limitations measured in benchmarks directly manifest as usability and confidence issues in practice. Our evaluation focuses on three well-established vulnerability classes where existing tools offer measurable performance. We acknowledge that emerging threats such as flash-loan and oracle-manipulation attacks remain underexplored due to limited analyzer support and low developer familiarity. As detection capabilities mature, expanding benchmarks to these modern attack vectors represents an important direction for future research.

From these results, we identify clear priorities for future research and tool development. In the short term, analyzers should focus on reducing false positives, improving explanation quality, and optimizing runtime to fit developer workflows. Longer-term directions include integrating hybrid analysis techniques, and expanding detection coverage to emerging attack vectors such as flash-loan and oracle manipulation. We also call for community collaboration on shared benchmarks and open datasets to ensure reproducible progress.